\documentclass[fleqn,12pt,twoside,english]{article}
\usepackage{espcrc1}
\usepackage{graphicx}
\usepackage{subfigure}
\usepackage{latexsym}
\usepackage{babel}

\newcommand{\AmS}{{\protect\the\textfont2
  A\kern-.1667em\lower.5ex\hbox{M}\kern-.125emS}}

\title{ {\bf A neutron star model in the nonlinear Relativistic
        Mean-Field Theory}}

\author{Ilona Bednarek
        \address{\it {Department of Astrophysics and
                      Cosmology, Institute of Physics,
                      University of Silesia, Katowice, Poland.}}
        \thanks{bednarek@us.edu.pl},
        Ryszard Ma\'{n}ka
	\thanks{manka@us.edu.pl}
       }

\begin{document}
\maketitle

\begin{abstract}
The neutron star parameters in the model extended
by the inclusion of $\delta$ meson and additional nonlinear vector
meson interactions are studied.
\end{abstract}

\section*{The relativistic nonlinear mean field model}

The aim of this paper is to study the influence of the additional
$\delta $ meson and nonlinear vector meson interactions on neutron
star parameters. The considered model comprises: nucleons, mesons
and leptons. The starting point is the construction of the effective
Lagrangian intended for application to the model described 
\begin{eqnarray}
\mathcal{L} & =\frac{1}{2}\partial _{\mu }\varphi \partial ^{\mu }\varphi -U(\varphi )-\frac{1}{4}\Omega _{\mu \nu }\Omega ^{\mu \nu }+\frac{1}{2}M_{\omega }^{2}\omega _{\mu }\omega ^{\mu }+\frac{1}{4}c_{3}(\omega _{\mu }\omega ^{\mu })^{2}+ & \label{lag}\nonumber \\
 & -\frac{1}{4}R_{\mu \nu }^{a}R^{a\mu \nu }+\frac{1}{2}M_{\rho }^{2}b_{\mu }^{a}b^{a\mu }+\frac{1}{4!}g_{\rho }^{4}\zeta (b_{\mu }^{a}b^{a\mu })^{2}+\frac{1}{2}\partial _{\mu }\delta ^{a}\partial ^{\mu }\delta ^{a}-\frac{1}{2}M_{\delta }^{2}\delta ^{a}\delta ^{a} & \nonumber \\
 & +(g_{\rho }g_{\omega })^{2}\Lambda _{v}b_{\mu }^{a}b^{a\mu }\omega _{\mu }\omega ^{\mu }+(g_{\rho }g_{s})^{2}\Lambda _{4}b_{\mu }^{a}b^{a\mu }\varphi ^{2}+ & \\
 & i\overline{\psi }\gamma ^{\mu }D_{\mu }\psi -\overline{\psi }(M-g_{s}\varphi -I_{3N}g_{\delta }\tau ^{a}\delta ^{a})\psi . & \nonumber 
\end{eqnarray}
The scalar sector consists of two isoscalar ($\sigma $ and $\omega $)
and two isovector ($\rho $ and $\delta $) mesons, denoted by $\varphi $,
$\omega _{\mu }$, $b_{\mu }^{a}$ and $\delta ^{a}$ respectively.
Meson $\delta $ has to be included as the model describes highly
asymmetric neutron star matter. $R_{\mu \nu }^{a}$, $\Omega _{\mu \nu }$,
$F_{\mu \nu }$ are the field tensors, the covariant derivative is
given by $D_{\mu }=\partial _{\mu }+1/2ig_{\rho }b_{\mu }^{a}\tau ^{a}+ig_{\omega }\omega _{\mu }$.
The potential function has the form 
\begin{equation}
U(\varphi )=1/2m_{s}^{2}\varphi ^{2}+1/3g_{2}\varphi ^{3}+1/4g_{3}\varphi ^{4}.
\end{equation}
The nucleon masses are denoted by $M_{N}$ whereas $m_{s}$, $M_{\omega }$,
$M_{\rho }$ and $M_{\delta }$ are masses assigned to the meson fields.\\
 The parameters entering the Lagrangian function (\ref{lag}) have
been chosen to reproduce the equilibrium properties of symmetric nuclear
matter. As it is usually assumed in quantum hadrodynamics \cite{walecka}
the mean field approximation is adopted and for the ground state of
homogeneous infinite matter, quantum fields operators are replaced
by their classical expectation values. Thus mesonic fields can be
separated into classical mean field values and quantum fluctuations
which are not included in the ground state: 
\begin{eqnarray}
\begin{array}{ll}
 \sigma =\overline{\sigma }+s & \delta ^{a}=\overline{\delta }^{a}+d\delta ^{3a}\\
 \omega _{\mu }=\overline{\omega }_{\mu }+w_{0}\delta _{\mu 0} & b_{\mu }^{a}=\overline{b}_{\mu }^{a}+r_{0}\delta _{\mu 0}\delta ^{3a}\end{array}
 &  & 
\end{eqnarray}
 The field equations derived from the Lagrangian function at the mean
field level are 
\begin{equation}
(m_{s}^{2}-2(g_{\rho }g_{s})^{2}\Lambda _{4}r_{0}^{2})s+g_{3}s^{2}+g_{4}s^{2}=\sum _{N}g_{s}M_{eff,N,k_{F}}^{2}S(M_{eff,N},k_{F,N})
\end{equation}
\begin{equation}
(m_{\omega }^{2}+2(g_{\rho }g_{\omega })^{2}\Lambda _{v}r_{0}^{2})w_{0}+c_{3}w_{0}^{3}=n_{B}
\end{equation}
\begin{equation}
(m_{\rho }^{2}+2(g_{\rho }g_{s})^{2}\Lambda _{4}s^{2})r_{0}+2(g_{\rho }g_{\omega })^{2}\Lambda _{v}r_{0}w_{0}^{2}+\frac{1}{6}g_{\rho }^{4}\zeta r_{0}^{3}=g_{\rho }I_{3N}n_{B}
\end{equation}
\begin{equation}
m_{\delta }^{2}d^{3}=\sum _{N}g_{\delta N}I_{3N}S(M_{eff,N},k_{F,N}).
\end{equation}
 Their forms indicate the in-medium modification of meson and nucleon
masses \cite{piek},\cite{ma:ef}. The function $S(M_{eff,N},k_{F,N})$
is expressed with the use of Fermi integrals 
\begin{equation}
S(M_{eff,N},k_{F,N})=\frac{2J_{N}+1}{2\pi ^{2}}\int _{0}^{k_{FN}}\frac{M_{N,eff}}{\sqrt{k^{2}+M_{N,eff}}}k^{2}dk
\end{equation}
where $J_{N}$ and $I_{3N}$ are the spin and isospin projection,
$k_{F,N}$ is the Fermi momentum of species N ($N=n,p$), $n_{B}$
denotes the baryon density. The effective nucleon mass which follows
from the Dirac equation has the form 
\begin{equation}
M_{N,eff}=M_{N}-g_{s}s+I_{3N}g_{\delta }d
\end{equation}
\centerline{Table 1}

\vspace*{0.15cm}

\centerline{Parameters of this model}

\vspace*{0.05cm}

\begin{center}\begin{tabular}{|c|c|c|c|c|c|}
\hline 
parameter set&
$g_{\delta }$&
$g_{\rho }$&
$\zeta $&
$\Lambda _{v}$&
 $\Lambda _{4}$\\
\hline
TM1&
 9.264 &
 0 &
 0. &
 0 &
 0 \\
\hline
GM3 &
 10.5&
 5.25 &
 0 &
 0 &
 0 \\
\hline
nonl. TM1 &
 3.5 &
 10.9 &
 0.5 &
0.008 &
 0.001  \\
\hline
\end{tabular}
\end{center}
The main effect of the inclusion of $\delta $ meson and nonlinear
vector meson interactions becomes evident studying properties of neutron
star matter, especially the form of the energy density $\varepsilon $
and the equation of state. These two equations include contributions
coming from meson, lepton and nucleon fields and are given in the
following forms 
\begin{eqnarray}
P & = & \frac{1}{2}m_{\rho }r_{0}^{2}+\frac{1}{2}m_{\omega }w_{0}^{2}+\frac{1}{4}c_{3}w_{0}^{4}-\frac{1}{2}m_{\delta }d^{2}+\Lambda _{v}(g_{\rho }g_{\omega })^{2}r_{0}^{2}w_{0}^{2}\\
 & + & \Lambda _{4}(g_{\rho }g_{s})^{2}r_{0}^{2}s^{2}+\frac{1}{24}\zeta g_{\rho }^{4}r_{0}^{4}-U(s)+P_{N}+P_{L}\label{ps:ress}\nonumber 
\end{eqnarray}
\begin{eqnarray}
\varepsilon  & = & \frac{1}{2}m_{\rho }^{2}(r_{0})^{2}+\frac{1}{2}m_{\delta }d^{2}+3\Lambda _{v}(g_{\rho }g_{\omega })^{2}r_{0}^{2}w_{0}^{2}+\Lambda _{4}(g_{\rho }g_{s})^{2}r_{0}^{2}s^{2}\\
 & + & \frac{1}{2}m_{\omega }^{2}w_{0}^{2}+\frac{3}{4}c_{3}w_{0}^{4}+\frac{1}{8}\zeta g_{\rho }^{4}r_{0}^{4}+U(s)+\epsilon _{N}+\epsilon _{L}.\label{en:dens}\nonumber 
\end{eqnarray}

\section*{Discussion}

The obtained form of the equation of state (\ref{ps:ress}) serves
as an input for the Oppenheimer-Tolman-Volkoff equations. In the result
hydrostatically stable configurations are calculated and the mass-radius
relation can be drown. Fig.3 depicts some models of neutron stars
which have been constructed for chosen parameter sets. In this figure
ZM denotes the Zimanyj- Moszkowski model, QMF - the model for quark
stars and GM3-the neutron stars model. For the given sequences of
models the gravitational binding energy have been calculated and results
are shown in Fig. 4. Analyzing the gravitational binding energy one
can come to the conclusion that the configuration with $\delta $
meson is energetically favorable than the one without $\delta $ meson.
However, for higher densities better results are obtained for the
ZM and QMF models. Throughout the effective nucleon masses the meson
$\delta $ alters nucleon chemical potentials what realizes in characteristic
modification of the appearance, abundance and distributions of the
individual species in neutron star matter. The inclusion of $\delta $
meson causes considerable nucleon mass splitting. In Fig.1 results
are shown for chosen parameters marked as set III. (Table I). The
appearance of additional nonlinear vector meson interactions produces
also visible effects on the nucleon and lepton abundance. From Fig.3
one can see results obtained for the parameter set II.\\
\begin{figure}
\begin{minipage}[t]{80mm}
\includegraphics[width=8.5cm]{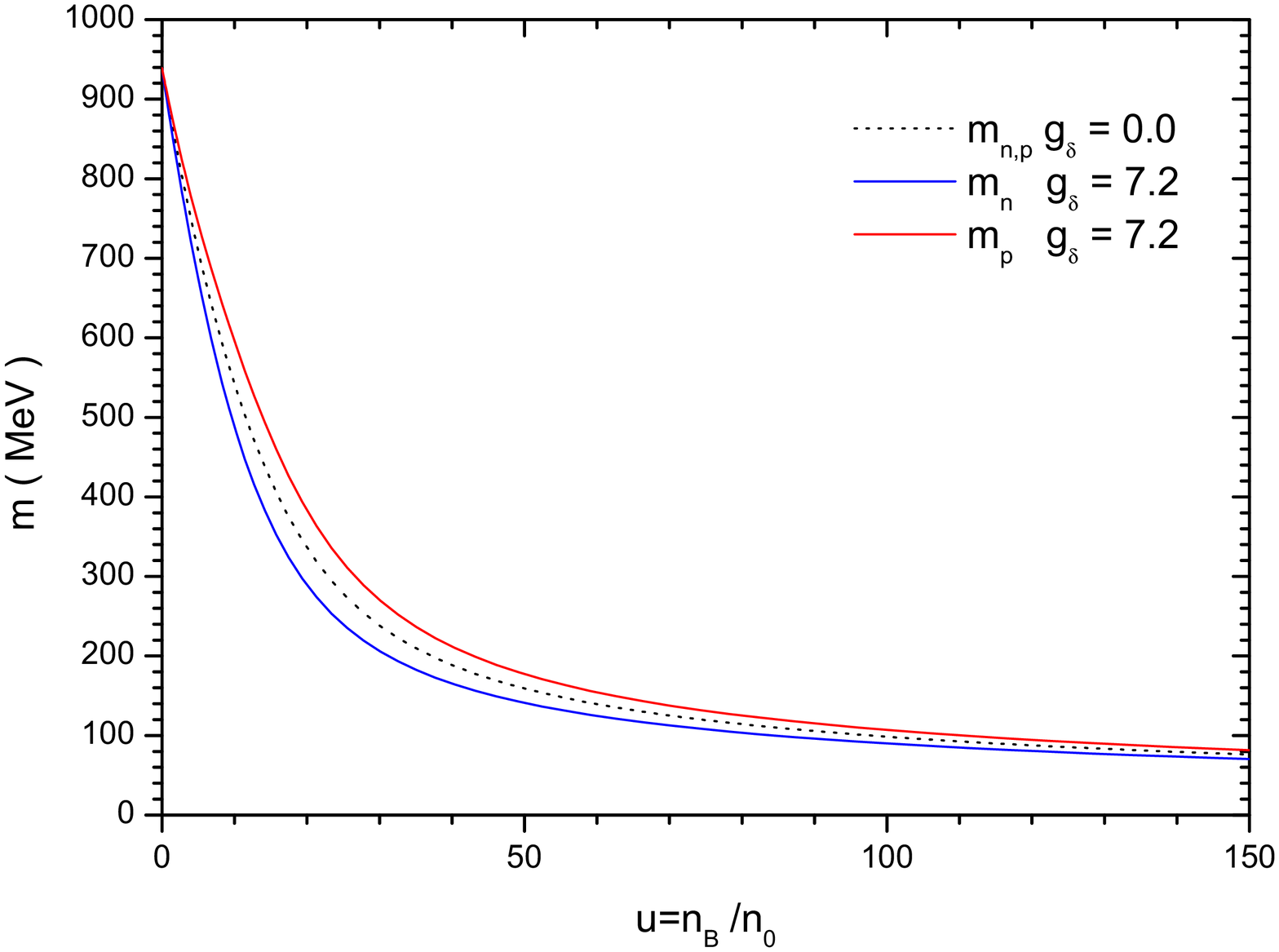}
\caption{The nucleon effective masses in the nonlinear RMF theory.}
\label{fig:rm}
\end{minipage}
\hspace{\fill}
\begin{minipage}[t]{75mm}
\includegraphics[width=8.5cm]{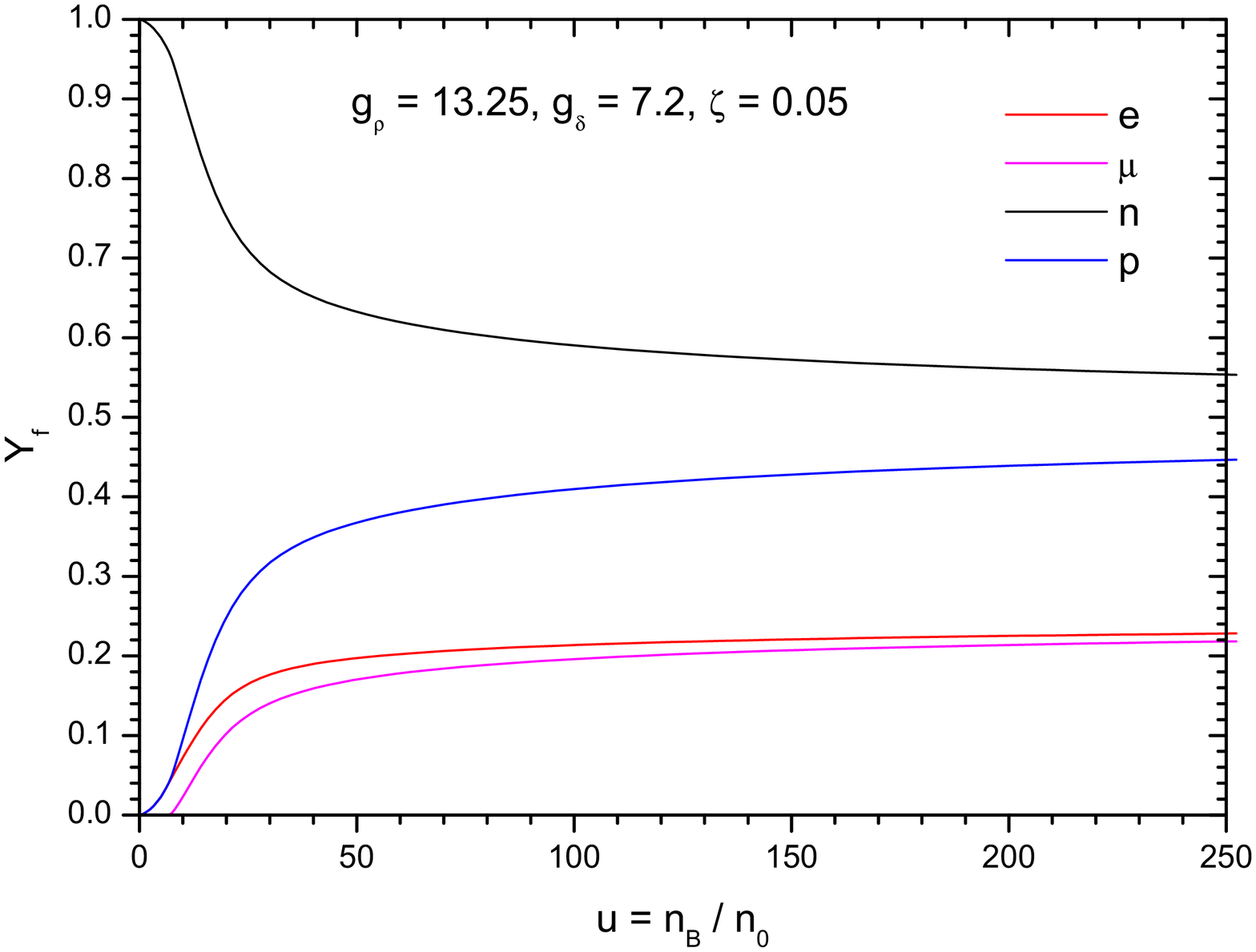}
\caption{Particle concentrations in the neutron star matter.}
\label{fig:yf2}
\end{minipage}
\end{figure}
\begin{figure}
\begin{minipage}[t]{80mm}
\includegraphics[width=8.5cm]{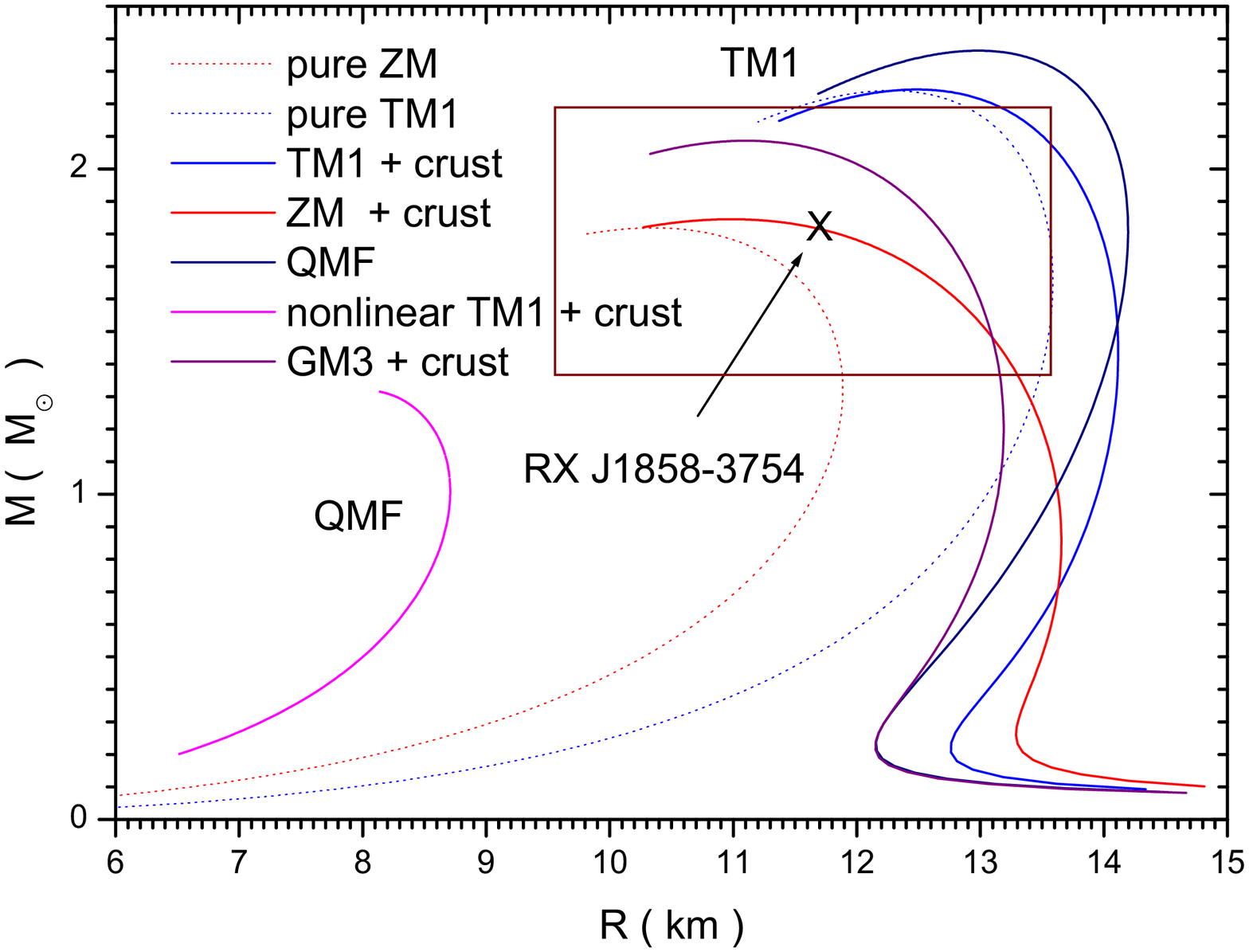}
\caption{The mass radius relations for  neutron stars in the RMF theory and for the quark
star in the QMF theory.}
\label{fig:rm1}
\end{minipage}
\hspace{\fill}
\begin{minipage}[t]{75mm}
\includegraphics[width=8.5cm]{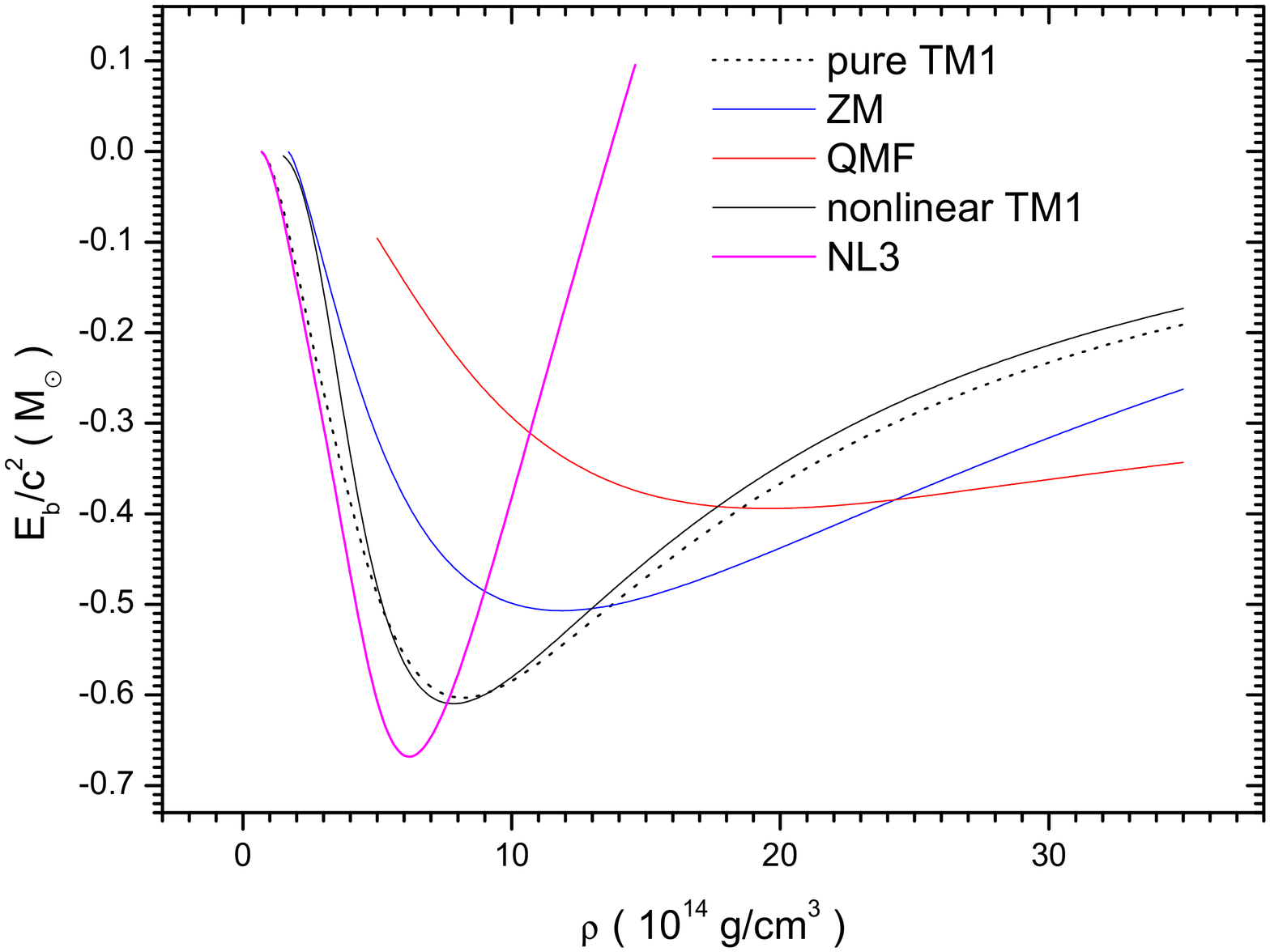}
\caption{Gravitational binding \(E_{g}\) energy as a function of central
density of neutron star and quark one.}
\label{fig:eberg}
\end{minipage}
\end{figure}
\newpage
\centerline{Table 2}
\vspace*{0.15cm}
\centerline{The neutron star properties in the nonlinear RMF model}
\vspace*{0.15cm}
\begin{center}
\begin{tabular}{|c|c|c|c|c|c|}
\hline 
parameter set&
$M_{max}(M_{\odot })$&
$R_{max}(km)$&
$M_{min}(M_{\odot })$&
$R_{min}(km)$&
$E_{g}$\\
\hline
\hline 
TM1 &
 2.20 &
13.8 &
0.2&
12.8&
 -0.617 \\
\hline 
GM3 &
 2.17 &
12.9 &
0.2&
12.2&
-0.602 \\
\hline 
nonl. TM1 &
 2.30 &
14.0&
0.2&
12.2&
-0.586 \\
\hline 
RX J185635-3754 &
 $1.7\pm 0.4$&
 $11.4\pm 2$&
 -  &
 -  &
 -  \\
\hline
\end{tabular}
\end{center}
The analyzed form of the equation of state leads to the conclusion
that the inclusion of $\delta $ meson and nonlinear vector meson
interactions causes the softening of the equation of state and in
the result the decrease of the neutron star parameters (Fig.3). Of
special interest is the case which includes $\delta $ meson, the
quartic $\rho $ meson term and nonlinear vector meson interactions
(set III in Table I). This case leads to the neutron star model with
the value of the maximum radius $R_{max}=14.0$ km. This result is
consistent with recent observations \cite{pr:RX} of RX J185635-3754.

\end{document}